\documentclass{iopart}

\usepackage{graphicx}
\usepackage{color}

\newcommand{\nc}{\newcommand}
\nc{\text}[1]{{\rm #1}}
\nc{\bra}[1]{\langle #1|}
\nc{\ket}[1]{|#1\rangle}
\nc{\braket}[1]{\left\langle #1 \right\rangle}
\nc{\dagg}{^{\dagger}}
\nc{\conj}{^{*}}
\nc{\dx}[1]{\, \mathrm{d} {#1} \,}
\nc{\Dx}[1]{\mathcal{D} {#1} \,}
\nc{\la}{\langle}
\nc{\ra}{\rangle}
\nc{\Id}{\mathbb{1}}
\nc{\eps}{\varepsilon}
\nc{\der}[2]{\frac{\mathrm{d} {#1}}{\mathrm{d} {#2}}}
\nc{\pder}[2]{\frac{\partial {#1}}{\partial {#2}}}
\nc{\bigO}{\mathcal{O}}
\nc{\half}{\frac{1}{2}}

\nc{\Eq}[1]{Equation (\ref{#1})}
\nc{\eq}[1]{equation (\ref{#1})}
\nc{\chap}[1]{Chapter~\ref{#1}}
\nc{\Sect}[1]{Section~\ref{#1}}
\nc{\sect}[1]{section~\ref{#1}}
\nc{\fig}[1]{figure~\ref{#1}}
\nc{\Fig}[1]{Figure~\ref{#1}}
\nc{\tabl}[1]{Table~\ref{#1}}
\nc{\app}[1]{Appendix~\ref{#1}}
\nc{\alg}[1]{\textcolor{blue}{#1}}

\begin{document}

%\title{Draft: Dissipative Dicke model with non-linear atom photon interaction}
\title[Generalized dissipative Dicke model]{Dissipative Dicke model with non-linear atom-photon interaction}

\author{A L Grimsmo$^{1,2}$ and A S Parkins$^1$}
\address{$^1$ Department of Physics, University of Auckland, Private Bag 92019, Auckland, New Zealand}
\address{$^2$ Department of Physics, The Norwegian University of Science and Technology, N-7491 Trondheim, Norway}
\ead{arne.grimsmo@ntnu.no}
\ead{s.parkins@auckland.ac.nz}

%\date{\today}

\begin{abstract}
We study a generalized Dicke model, as recently realized in an atomic quantum gas experiment, describing the collective interaction of $N$ two-level atoms with a single cavity mode. The model takes account of dissipation of the cavity field, and includes a non-linear atom-photon coupling, not present in the conventional Dicke model. We extend previous theoretical investigations of a semiclassical model by including all quantum effects and considering finite atom number $N$. Our results show good agreement between quantum expectation values and the semiclassical model as $N$ is increased, but also show exotic behaviour for the corresponding quantum state as the non-linear atom-photon coupling is varied.
\end{abstract}

% 42.50.Pq cavity qed
% 37.30.+i atoms in cavities
% 42.50.Ct Light - interaction with matter
% 05.70.Fh phasetransitions in stat mech and thermodynamics


\maketitle

\section{Introduction}

Fundamental models of many-body physics, originally developed as approximate theoretical descriptions of highly complex underlying systems, have in recent years been realized experimentally with an unprecedented degree of control thanks to advances in atomic physics and quantum optics. In particular, systems using ultracold atomic gases in optical lattices \cite{Bloch08} and systems based on cavity quantum electrodynamics (cavity QED) \cite{Baumann10} have provided effective realizations (or ``quantum simulations'' \cite{Buluta09,Cirac12}) of many-body models with a high degree of accuracy via exquisite experimental control and manipulation of atoms and electromagnetic fields. Such approaches also offer a flexible means of varying the effective model parameters, opening up exciting possibilities for exploring several frontiers of many-body physics and of strongly interacting quantum systems in general. 

Such realizations are particularly interesting in light of quantum phase transitions in many-body systems \cite{Sachdev07}, as the flexibility of the schemes allows ready access to different phases and to the boundary (critical) regions between them. In particular, transitions between different phases can, for example, be observed in real time by the slow variation of one or more controllable system parameters (see, e.g., \cite{Baumann10}). One can also consider more rapid ``quench'' dynamics by a sudden change of parameters, taking the system from one region of phase space to a distinctly different region over a short period of time (see, e.g., \cite{Baumann11} and references therein) and thereby inducing complicated, non-equilibrium dynamics. In addition, signatures of the critical behaviour are observable in the output channels from the systems; for example, in cavity QED systems, one typically finds dramatic changes in the properties of the output field of the cavity, such as the intensity and the photon counting statistics, and, for example, spontaneous symmetry breaking at a critical point may in fact be directly observable through homodyne detection of the field \cite{Baumann11}\footnote{One should note that observing spontaneous symmetry breaking is typically extremely hard, as any external fluctuations or imperfections might destroy the necessary symmetry, but recent results have proved that atomic quantum gas experiments provide a promising route to observing such phenomena \cite{Baumann11}.}. 

Another interesting aspect of such effective realizations is that they often offer generalizations of the original model, sometimes in terms of non-trivial extensions of the Hamiltonian of the system, and also due to the intrinsically open (dissipative) nature of atomic and cavity QED systems, which in itself can lead to new phenomena. One particular example of this, which will be our focus in this work, is the effective realization of the Dicke model quantum phase transition using a superfluid gas in an optical cavity, recently demonstrated in \cite{Baumann10}. Here, the two-level atoms of the original Dicke model are realized using resonant Raman transitions between a pair of discrete momentum states of a Bose-Einstein condensate (BEC). When comparing with the original Dicke model \cite{Dicke54,Hepp73}, this realization offers a number of generalizations, such as the non-negligible decay of the optical cavity field, independently tunable interaction strengths for rotating and counter-rotating terms in the atom-field interaction Hamiltonian, and a new non-linear coupling term between the cavity photon number and the collective atomic inversion. The implications of these generalizations have been studied theoretically in great detail in \cite{Keeling10,Bhaseen12} in the ``thermodynamic limit'' of a large number of atoms, where a semiclassical model can be employed. This analysis predicts an exceedingly rich dynamical phase diagram for the steady state of the system with fundamentally new behaviour and phases, such as a new superradiant phase, co-existence regions, and regimes with oscillatory long-time attractors. 

The theoretical results of \cite{Keeling10,Bhaseen12} focus on parameters appropriate to the experiment in \cite{Baumann10}. An analogous realization of the same model could also be based upon Raman transitions between a pair of electronic hyperfine ground states in alkali atoms, as first proposed and investigated theoretically in \cite{Dimer07}. This leads to an identical effective Dicke model, but the scheme potentially offers access to a broader range of the effective parameters, and alternative ways of tuning them in order to access different regions of phase space. Further investigation of this scheme in \cite{Grimsmo13}, albeit focussing on the single-atom case, suggests that rubidium atoms coupled to a high finesse optical cavity should enable a good realization of the model. A many-atom cavity QED setup as realized recently in \cite{Arnold12} is potentially of most immediate relevance to the current work.

In particular, in this work we study this effective Dicke model including all quantum effects, and considering finite numbers of atoms. How this quantum model connects with the results for the thermodynamic limit considered previously is one aspect of our analysis, but, to further complement the work in \cite{Keeling10,Bhaseen12}, we also consider somewhat different parameter regimes, motivated by the possibility of realizing the model using the scheme discussed in the previous paragraph. We will see that one can achieve good acccess to the distinct regions of parameter space, with parameter ranges believed feasible based on the results from \cite{Grimsmo13}.
%We will see that this model offers good access to the distinct regions of parameter space, with parameter ranges believed feasible based on the results from \cite{Grimsmo13}. 
Furthermore, our findings indicate that one can observe phase transitions between a range of distinctly different phases through the variation of a single model parameter.

\section{Quantum model and semiclassical approximation}

The generalized Dicke model that we consider describes the interaction of $N$ two-level atoms with a single mode of the electromagnetic field. The dynamics can be given in terms of a master equation for the density operator, $\rho$, of the atoms and cavity mode, as \cite{Dimer07}:
\begin{eqnarray}\label{eq:master}
\dot{\rho} = -i[H,\rho] + 2\kappa \mathcal{D}[a]\rho,
\end{eqnarray}
where
\begin{eqnarray}\label{eq:H_Dicke}
H =& \omega_0 J_z + \omega a\dagg a + \frac{g}{\sqrt{N}}\left(J_- + J_+\right)\left(a + a\dagg\right) \\
&+ \frac{U}{N} J_z a\dagg a \nonumber,
\end{eqnarray}
and $\mathcal{D}$ is a super-operator defined through $\mathcal{D}[O]\rho = O\rho O\dagg - 1/2 O\dagg O \rho - 1/2 \rho O\dagg O$, for any operator $O$. Parameters $\omega_0$ and $\omega$ are the atomic and cavity frequencies, respectively, $g$ is the linear interaction strength, and $U$ is a non-linear coupling constant. Operators $a$ and $a\dagg$ are the annihilation and creation operators for the cavity field, and $\{J_z, J_-, J_+\}$ are collective atomic operators satisfying angular momentum commutation relations: $[J_+,J_-] = 2J_z$, $[J_\pm,J_z] = \mp J_\pm$. \Eq{eq:master} reproduces the conventional equilibrium Dicke model if $U = \kappa = 0$.

This generalized Dicke model has been studied theoretically in \cite{Keeling10,Bhaseen12} in the thermodynamic limit, $N \rightarrow\infty$. In this limit one can derive a closed set of equations for the expectation values 
\begin{eqnarray}
  \alpha \equiv& \frac{\braket{a}}{\sqrt{N}} , \nonumber \\
   \beta \equiv& \frac{\braket{J_-}}{N} , \\
  \gamma \equiv& \frac{\braket{J_z}}{N} \nonumber.
 \end{eqnarray}
Using \eq{eq:master} and assuming factorization of operator products, i.e., $\braket{J_{z,\pm}(a+a\dagg)} = \braket{J_{z,\pm}}\braket{a+a\dagg}, \braket{J_-a\dagg a} = \braket{J_-}\braket{a\dagg a}=\braket{J_-}|\braket{a}|^2$, one can arrive at the following semiclassical equations of motion:
\begin{eqnarray}
\dot{\alpha} =& -i\left(\omega -i\kappa + U\gamma\right)\alpha - ig(\beta+\beta\conj) , \nonumber \\
\dot{\beta} =& -i\left(\omega_0 + U|\alpha|^2\right)\beta + 2ig\left(\alpha+\alpha\conj\right)\gamma , \label{eq:semiclass} \\
\dot{\gamma} =& ig(\alpha+\alpha\conj)(\beta-\beta\conj). \nonumber
\end{eqnarray}
We note that both the quantum and semiclassical models conserve the total length of the spin. Physical initial conditions of course require that $\braket{J_x}^2 + \braket{J_y}^2 + \braket{J_z}^2 \le N^2/4$, where $J_{\pm} = J_x \pm iJ_y$. In \cite{Keeling10,Bhaseen12} dynamical phase diagrams were mapped out by considering the long time attractors of \eq{eq:semiclass}. The semiclassical analysis predicts an exceedingly rich phase diagram, where the generalizations due to cavity decay and the non-linear coupling ($U$) leads to fundamentally new behaviour and phases, including a new superradiant phase, coexistence regions, and regimes of persistent oscillations for $\alpha$ and $\beta$. Since one of our goals is to compare the semiclassical predictions with fully quantum results for finite $N$, we first summarize briefly some of the main results from \cite{Keeling10,Bhaseen12}.

The steady states of \eq{eq:semiclass} can be found analytically by requiring $\dot{\alpha} = \dot{\beta} = \dot{\gamma} = 0$. One finds that the semiclassical normal phase, $\{\alpha = \beta = 0$, $\gamma = -1/2\}$, and inverted phase $\{\alpha = \beta = 0$, $\gamma = +1/2\}$ are always valid fixed points. In addition, two superradiant phases, referred to as superradiant A (SRA) and superradiant B (SRB), are possible. The SRA phase refers to solutions where $\braket{J_y}/N = i/2(\beta - \beta\conj) = 0$ and the SRB phase to solutions where $\omega_0 + U|\alpha|^2 = 0$. We will assume $\omega_0 \ge 0$ throughout this paper, so that the latter condition is only possible for $U \le 0$. The onset of superradiance from the normal/inverted phase was identified at critical values of the coupling constant $g$. The SRA phase comes into existence for $g \ge g_A^{\pm}$ where $+$ refers to the onset from the normal phase, $-$ from the inverted phase, and
\begin{eqnarray}\label{eq:g_ca}
  g_A^{\pm} = \sqrt{\mp \frac{\omega_0}{4(\omega \pm U/2)}(\kappa^2 + (\omega \pm U/2)^2)}.
\end{eqnarray}
The steady state values in this phase are
\numparts
\begin{eqnarray}
\gamma^{A\pm} &= -\frac{\omega}{U} \pm \sqrt{\frac{g^2(4\omega^2-U^2)-\omega_0 U \kappa^2}{U^2(\omega_0 U + 4g^2)}} , \label{eq:Jz_sc_A_pm} \\
%|\alpha^{A\pm}| &= |2g\sqrt{\frac{1/4-(\gamma^{A\pm})^2}{\omega+U \gamma^{A\pm} -i\kappa}}|.
\alpha^{A+\pm} &= \pm2g\sqrt{\frac{1/4-(\gamma^{A+})^2}{\omega+U \gamma^{A+} -i\kappa}} , \\
\alpha^{A-\pm} &= \pm2g\sqrt{\frac{1/4-(\gamma^{A-})^2}{\omega+U \gamma^{A-} -i\kappa}}.
\end{eqnarray}
\endnumparts
Similarly, the SRB phase comes into existence if $g \ge g_B$ where
\begin{eqnarray}
  g_B = \kappa \sqrt{\frac{\omega_0 U}{4(\omega^2 - (U/2)^2)}}, \qquad U \le 0,
\end{eqnarray}
and the steady state values are
\numparts
\begin{eqnarray}
\gamma^{B} &= -\frac{\omega}{U} \label{eq:Jz_sc_B} , \\
\alpha^{B\pm} &= \pm i\sqrt{\frac{-\omega_0}{U}}.
%\alpha^{B\pm} &= \pm i\sqrt{\frac{-\omega_0}{U}}.
\end{eqnarray}
\endnumparts
Since these results imply that multiple fixed points exists at any given point in parameter space, a further stability analysis of the different steady states is necessary to determine the phase diagram. However, one can immediately draw a few conclusions based on the above results. Physical solutions require $\gamma \le 1/2$, and of course $\gamma$ must be real. This means that for the two roots of the SRA solution $\gamma^{A\pm}$, the $-$ solution can only exist for $U < -2\omega$, and the $+$ solution only for $U < 2\omega$. Similarly, the SRB solution can only be valid if $U < -2\omega$ (recall that we assume $\omega_0 \ge 0$). The two critical points for $U$ at $U = \pm 2\omega$ are found to mark the change in stability for the normal and inverted phases, the inverted phase being stable only if $U < -2\omega$ and the normal phase only if $U < 2\omega$. We do not reproduce the semiclassical stability analysis here, but refer to \cite{Bhaseen12} for details. We note in particular that the region $U \ge 2\omega$ has no stable semiclassical fixed points. It does yield a steady state solution for $\gamma$ given by \cite{Bhaseen12}
\begin{eqnarray}
\gamma^{U\ge2\omega} = -\frac{\omega}{U},
\end{eqnarray}
as in the SRB phase, but the region has periodic orbits as long time attractors for $\alpha$ and $\beta$. 

\begin{figure}
  \centering
  \includegraphics{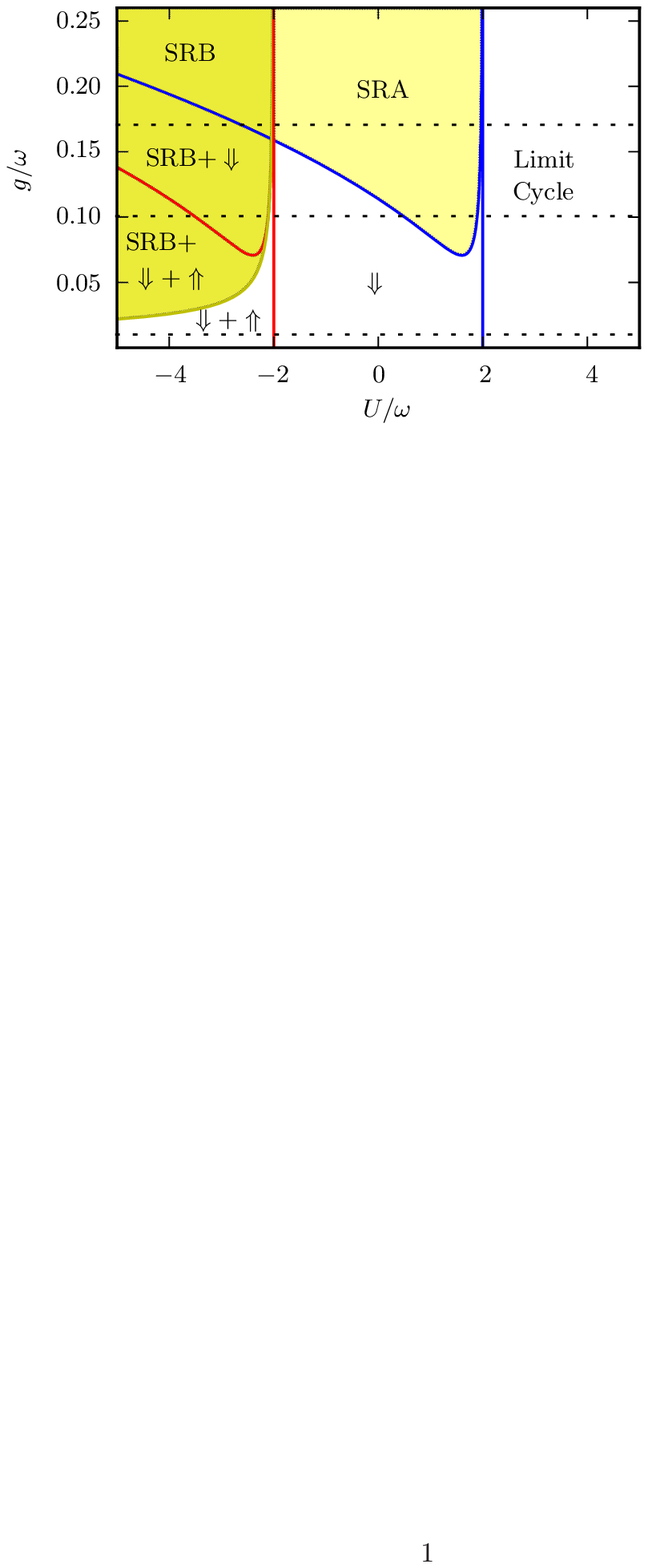}
  \caption{\label{fig:phasedia} Schematic phase diagram based on the semiclassical approximation. The two superradiant phases are denoted SRA and SRB. $\Downarrow$ refers to the normal phase $\{\alpha,\beta,\gamma\} = \{0,0,-1/2\}$, $\Uparrow$ to the inverted phase $\{\alpha,\beta,\gamma\} = \{0,0,+1/2\}$, and ``Limit Cycle'' to the region $U>2\omega$, with periodic orbits as the semiclassical long time attractors for $\alpha$ and $\beta$. The $+$ sign indicates co-existence phases. The dashed lines show three horizontal cuts, at $g/(2\pi) = \{ 0.01,0.1,0.17\}$ MHz, of the phase diagram that we will consider in the following sections.}
\end{figure}

The results of the semiclassical steady state analysis can be summarized in a phase diagram for $g$ and $U$. This is shown in \fig{fig:phasedia} for parameters $\omega/(2\pi) = 1.0$ MHz, $\omega_0/(2\pi) = 0.05$ MHz and $\kappa/(2\pi) = 0.2$ MHz (the specific values of which are motivated by \cite{Grimsmo13}). These results are analogous to the phase diagram in Figure 4 of Ref. \cite{Keeling10}. In particular, this diagram suggests that there is a possibility of taking the system through a whole range of different phases upon the variation of a single parameter, $U$. %The negative root of the SRA solution, \eq{eq:Jz_sc_A_pm}, does not appear in the phase diagram when $\omega > 0$, which will be the case in all our results to follow.

\section{Numerical results for finite $N$}

In this section, we present numerical solutions of \eq{eq:master} as the parameter $U$ is varied, and connect this to the semiclassical predictions discussed above. We focus our interest primarily on the steady state of \eq{eq:master}, numerically approximated by an inverse power method iterated until an absolute tolerance of $10^{-6}$ is achieved \cite{Johansson12}. Alternatively, steady states could be approximated by integrating \eq{eq:master} for long times, but this approach is problematic due to the exceedingly long time-scales involved in certain regions of parameter space, as pointed out in \cite{Bhaseen12} (see also \sect{sec:corrspectra} below). The cavity mode Hilbert space was approximated with a truncated Fock space, with up to nine states for the largest value of $g$ considered.

We focus here on the parameters used in \fig{fig:phasedia}, i.e., $\omega/(2\pi) = 1.0$ MHz, $\omega_0/(2\pi) = 0.05$ MHz, $\kappa/(2\pi) = 0.2$ MHz, $U$ in the range $-10.0$ MHz $\le U/(2\pi) \le 10.0$ MHz, and $g/(2\pi)$ up to 0.17 MHz. As for the number of atoms, we vary $N$ up to a maximum value of $N=90$. %This turns out to be sufficient to clearly see the various phase transitions predicted in \fig{fig:phasedia}.

A first observation that we would like to draw attention to, however, is that in regions of parameter space where the semiclassical analysis predicts co-existence phases, the size of the basin of attraction typically varies greatly for the individual fixed points. If there are two, or more, fixed points with appreciably large basins of attraction compared to the characteristic size of quantum fluctuations, one would expect the steady state of the ensemble average dynamics given by \eq{eq:master} to consist of a mixture of the solutions centered around the semiclassical fixed points. But, the semiclassical analysis does not predict the weighting of different fixed points, and we find this to in general depend on $N$ and the other parameters. If a fixed point which is stable in the asymptotic limit, $N\to\infty$, has a basin of attraction that is too small compared to quantum fluctuations for finite $N$, this fixed point will not contribute to the steady state of the quantum dynamics. We will often see this in our numerical results to follow. Another observation is that, for finite $N$, the normal ($\Downarrow$) and inverted ($\Uparrow$) phases, $\ket{0,\Downarrow/\Uparrow} \equiv \ket{0}_\text{cav}\otimes\ket{J_z=\mp N/2}_\text{spin}$, are in fact never the exact steady states of the quantum dynamics, as long as $g>0$. Indeed, even in regions of parameter space where the semiclassical analysis predicts one of these phases as the unique stable fixed point, one finds a small, but non-zero, photon number in the cavity for finite $N$. This photon-production comes from the counter-rotating terms in \eq{eq:H_Dicke}, and might at first glance seem unphysical. We must here remember though, that \eq{eq:master} is an effective model that comes about via laser-pumped Raman transitions between distinct states of a multilevel system (such as an alkali atom or BEC), and the photon flux from the cavity can thus be explained by the energy input of the lasers involved in the effective realization. The normal and inverted phases are only approached in the limit $N\to\infty$.

With this in mind, we vary the parameter $U$ along the horizontal cuts, shown as dashed lines in \fig{fig:phasedia}, and compare the semiclassical predictions to steady state expectation values given by \eq{eq:master}. We also find it useful, as a means of shedding further light on the various phases and phase transitions, to look at how the reduced density operator for the cavity field, $\rho_\text{cav}$, and the collective spin, $\rho_\text{spin}$, change. This can be visualized with, for example, the Wigner function and the atomic $Q$-function, respectively. Here the Wigner function for the cavity field is defined as
\begin{eqnarray}
  W(\alpha) = \frac{2}{\pi}\tr\left[D\dagg(\alpha)\rho_\text{cav}D(\alpha)(-1)^{a\dagg a}\right],
\end{eqnarray}
where $\alpha$ is a complex number and $D(\alpha) = \exp(\alpha a\dagg - \alpha\conj a)$ is the coherent displacement operator. The atomic $Q$-function is defined by
\begin{eqnarray}
  Q(\eta) = \braket{\eta|\rho_\text{spin}|\eta},
\end{eqnarray}
where $\ket{\eta}$ is the spin-coherent state,
\begin{eqnarray}
  \ket{\eta} = (1+|\eta|^2)^{-N/2} \sum_{n=0}^{N} {N \choose n}^{1/2} \eta^n \ket{n,N/2}.
\end{eqnarray}
Here, $\eta = \exp(i\phi)\tan(\theta/2)$, with $\theta$ and $\phi$ spherical coordinates, and $\{\ket{n,N/2}\}$ are the symmetric Dicke states (see, e.g., \cite{Stockton03}).

\subsection{Low $g$ behaviour: $g/(2\pi) = 0.01$ {\rm MHz}}

\begin{figure}
  \centering
  \includegraphics{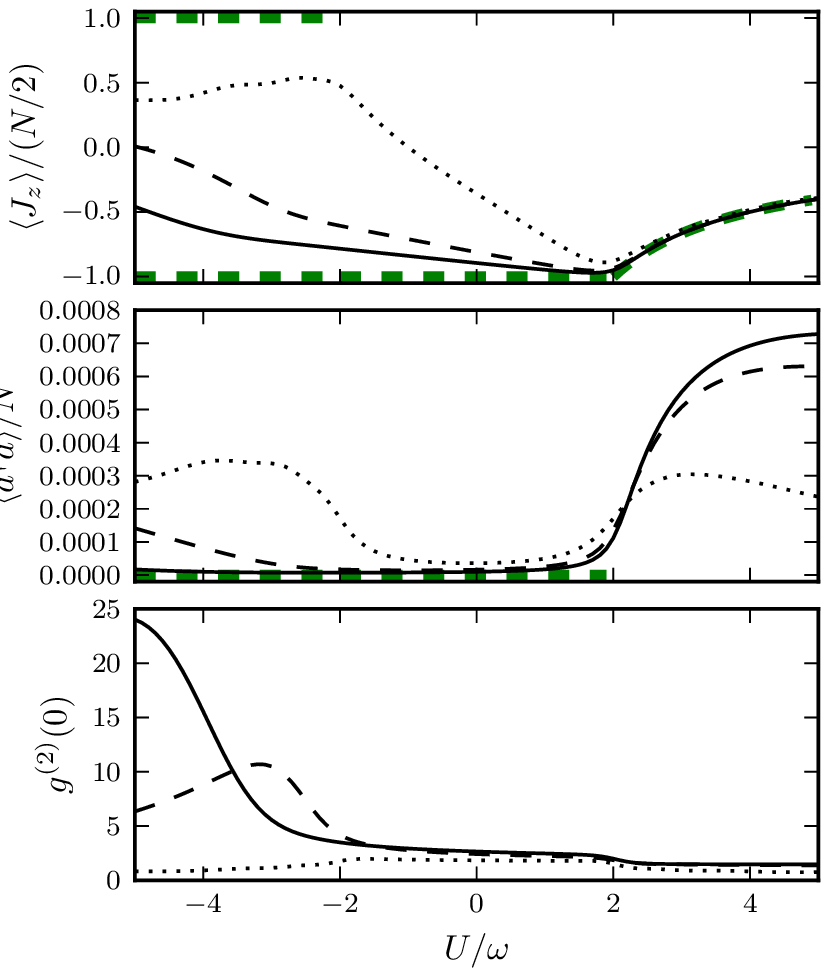}
  \caption{\label{fig:output_g0.01} Steady state expectation values, $\braket{J_z}$, $\braket{a^\dag a}$ and $g^{(2)}(0)$, as $U$ is varied along the lowest horizontal cut in \fig{fig:phasedia}: $\{\omega_0,\omega,\kappa,g\} = \{0.05,1.0,0.2,0.01\}\,2\pi\times\text{MHz}$: $N=10$ (dotted), $N=50$ (dashed), $N=90$ (solid). Semiclassical solutions are included as dashed green lines for comparison.}
\end{figure}
\begin{figure}
  \centering
  \includegraphics{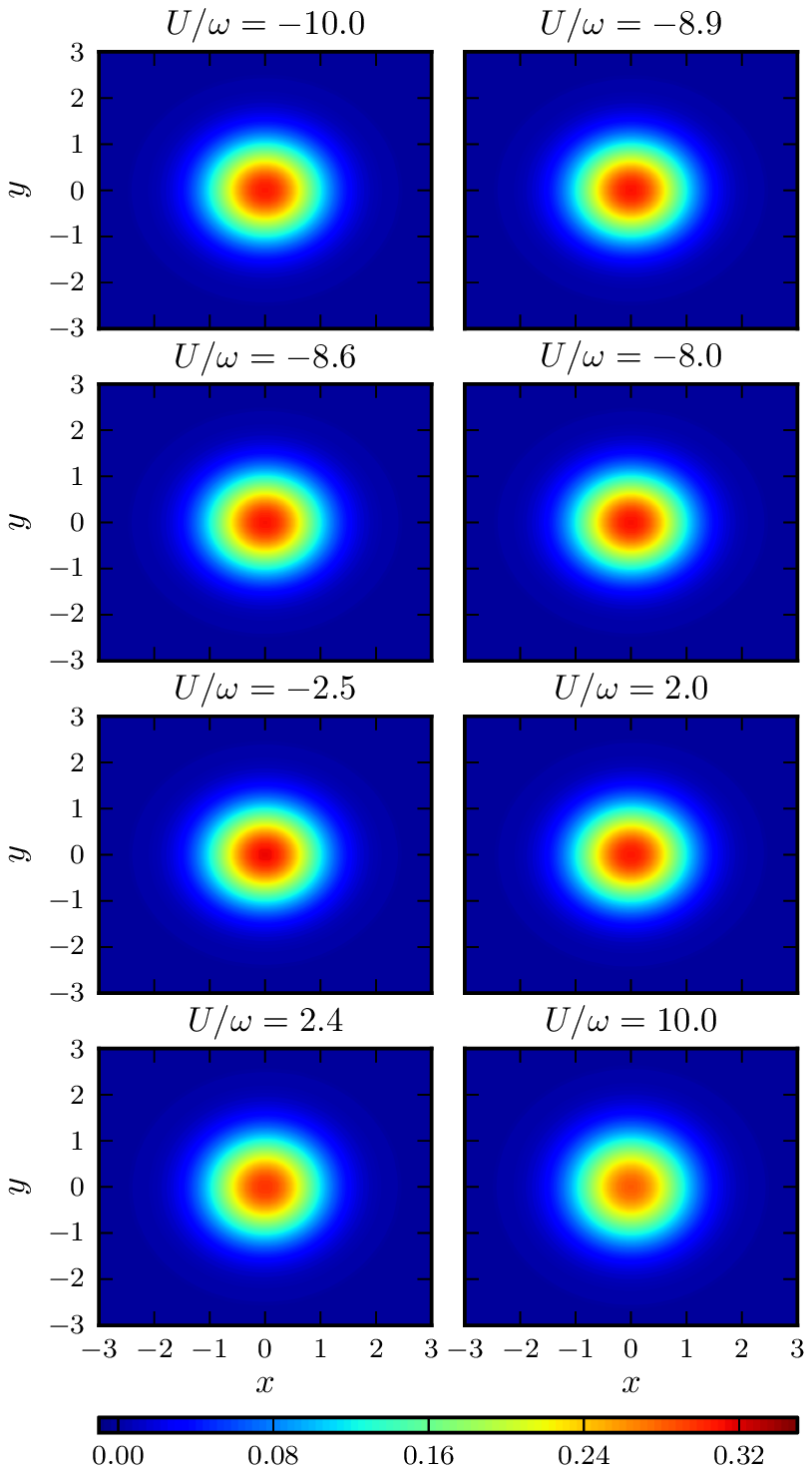}
  \caption{\label{fig:wigner_g0.01} Contour plots of the Wigner function $W(\alpha)$, with $\alpha = (x+iy)/\sqrt{2}$, for a series of values of $U$ along the lowest horizontal cut in \fig{fig:phasedia}: $\{\omega_0,\omega,\kappa,g\} = \{0.05,1.0,0.2,0.01\}\,2\pi\times\text{MHz}$, $N=90$.}
\end{figure}
\begin{figure}
  \centering
  \includegraphics{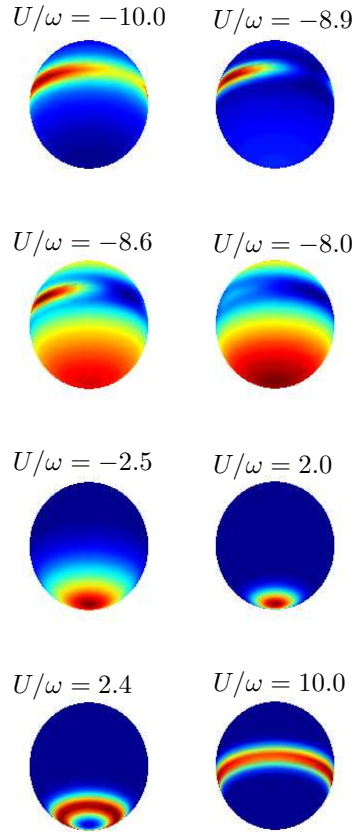}
  \caption{\label{fig:qfunc_g0.01} Steady state atomic $Q$-function, $Q(\theta,\phi)$, plotted on the Bloch sphere, for a series of values of $U$ along the lowest horizontal cut in \fig{fig:phasedia}: $\{\omega_0,\omega,\kappa,g\} = \{0.05,1.0,0.2,0.01\}\,2\pi\times\text{MHz}$, $N=90$. The colour scheme goes from blue to red, indicating low to high concentration, respectively. Note that the individual subplots are not on the same color scale.}
\end{figure}

We start our discussion on the finite $N$ results by considering the behaviour as $U$ is varied at low $g$; in particular, for $g/(2\pi) = 0.01$ MHz. This corresponds to the lowest dashed line in \fig{fig:phasedia}. There are no superradiant phases in this region of parameter space, but the semiclassical analysis suggests a transition from a co-existent $\Downarrow + \Uparrow$ phase for $U < -2\omega$ to a $\Downarrow$ phase for $-2\omega < U < 2\omega$. Furthermore, there is a transition to the ``limit cycle'' region for $U > 2\omega$, where the semiclassical solutions for $\alpha(t)$ and $\beta(t)$ oscillate. It is not obvious from the semiclassical analysis how these transitions might manifest themselves in the quantum state of the system, i.e., the steady state of the quantum master equation (\ref{eq:master}).

We first look, in \fig{fig:output_g0.01}, at the steady state expectation values (or ``order parameters'') $\braket{J_z}$ and $\braket{a\dagg a}$ as $U$ is varied along the cut $g/(2\pi) = 0.01$ MHz. We also show in the same figure the intensity correlation function of the cavity field, $g^{(2)}(0) = \braket{a\dagg a\dagg a a}/\braket{a\dagg a}^2$, in the steady state. We compare results for $N=10$, $50$ and $90$ atoms. The semiclassical solutions, as given in the previous section, are shown as green dashed lines wherever they apply, for comparison. As expected, the photon number in the cavity is small throughout, but there is a sharp increase at $U = 2\omega$. A pronounced increase is also seen in the collective atomic inversion, $\braket{J_z}$, at this point, with a shape that is characteristic of a second order phase transition. 

As for the region $U < -2\omega$, the behaviour of the system is seen to vary greatly with $N$, and, for small $N$, also with $U$. Regarding the dependence on $N$, we point out that the energies of the states $\ket{0,\Downarrow}$ and $\ket{1,\Uparrow}$ are degenerate at $U = -2\omega_0 N$. The same is true for $\ket{0,\Uparrow}$ and $\ket{1,\Downarrow}$. Degeneracies in the Hamiltonian of the ground state and an excited state at a critical point is the trademark of equilibrium quantum phase transitions \cite{Sachdev07}. We are considering a dynamical steady state of a quantum master equation, but with a degeneracy in the spectrum at a value of $U \sim N$, certainly a significant dependence on $N$ is not surprising. Note that this is not predicted in the thermodynamic limit, as the degeneracy is pushed out to $U = -\infty$. One might wonder if there are any sharp changes to the steady state at this point of degeneracy in the spectrum. The expectation values in \fig{fig:output_g0.01} do not indicate this. However, there is, in fact, a sharp transition in the quantum state of the \emph{spin} as $U$ crosses $-2\omega_0 N$. This can be visualized using the atomic $Q$-function, as defined in the previous section and considered below.

We study the reduced field and atomic states by looking at the cavity mode Wigner function and the atomic $Q$-function. Contour plots of the Wigner function for a series of values of $U$, with $N=90$, are shown in \fig{fig:wigner_g0.1}. We see that the state of the field stays close to the vacuum for all values of $U$, with a symmetric Wigner function concentrated around the origin. The atomic $Q$-function is plotted on the Bloch sphere for the same values of $U$, and $N=90$, in \fig{fig:qfunc_g0.01}. Here, significant variation in the atomic state can be seen. Starting out as a ring shape for large, negative $U$, there is an increasing concentration around two points, approaching a two-peak structure as $U$ is close to $U=-9.0\omega$. As $U$ crosses this point, the $Q$-function changes into a new two-peak structure, with a concentration around the north and south pole---this is what we would expect for a $\Downarrow + \Uparrow$ co-existent phase. As $U$ is increased further towards $U=-2.0\omega$, the concentration shifts more and more towards the south pole, i.e., the $\Downarrow$ phase, and at $U = -2.5\omega$, no concentration around the inverted phase can be seen. The atomic state stays in close approximation to the $\Downarrow$ phase throughout the interval $-2\omega < U < 2\omega$, but then there is a new sudden change around $U=2.0\omega$, where a ring like structure forms, and moves up towards the equator as $U$ grows. This behaviour for the quantum state corresponds to the semiclassical predictions of a steady state value for $\braket{J_z}$, $\braket{J_z}/N = -\omega/U$, and periodic orbits for $\braket{J_{\{x,y\}}}$. At large positive $U$, the state shares some similarities with the state at the corresponding negative $U$.

\subsection{$g/(2\pi) = 0.1$ {\rm MHz}}

\begin{figure}
  \centering
  \includegraphics{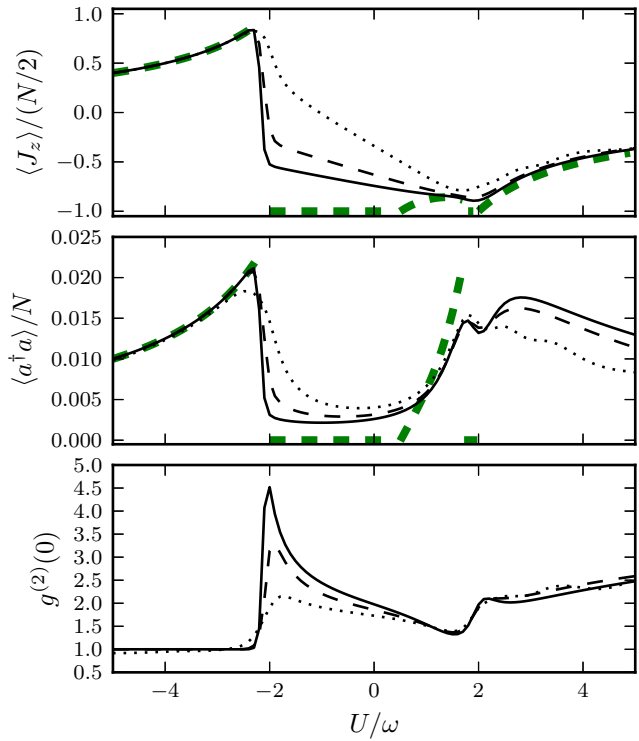}
  \caption{\label{fig:output_g0.1} Steady state expectation values, $\braket{J_z}$, $\braket{a^\dag a}$ and $g^{(2)}(0)$, as $U$ is varied along the middle horizontal cut in \fig{fig:phasedia}: $\{\omega_0,\omega,\kappa,g\} = \{0.05,1.0,0.2,0.1\} 2\pi\times\text{MHz}$: $N=10$ (dotted), $N=30$ (dashed), $N=50$ (solid). Semiclassical solutions are included as dashed green lines for comparison.}
\end{figure}
\begin{figure}
  \centering
  \includegraphics{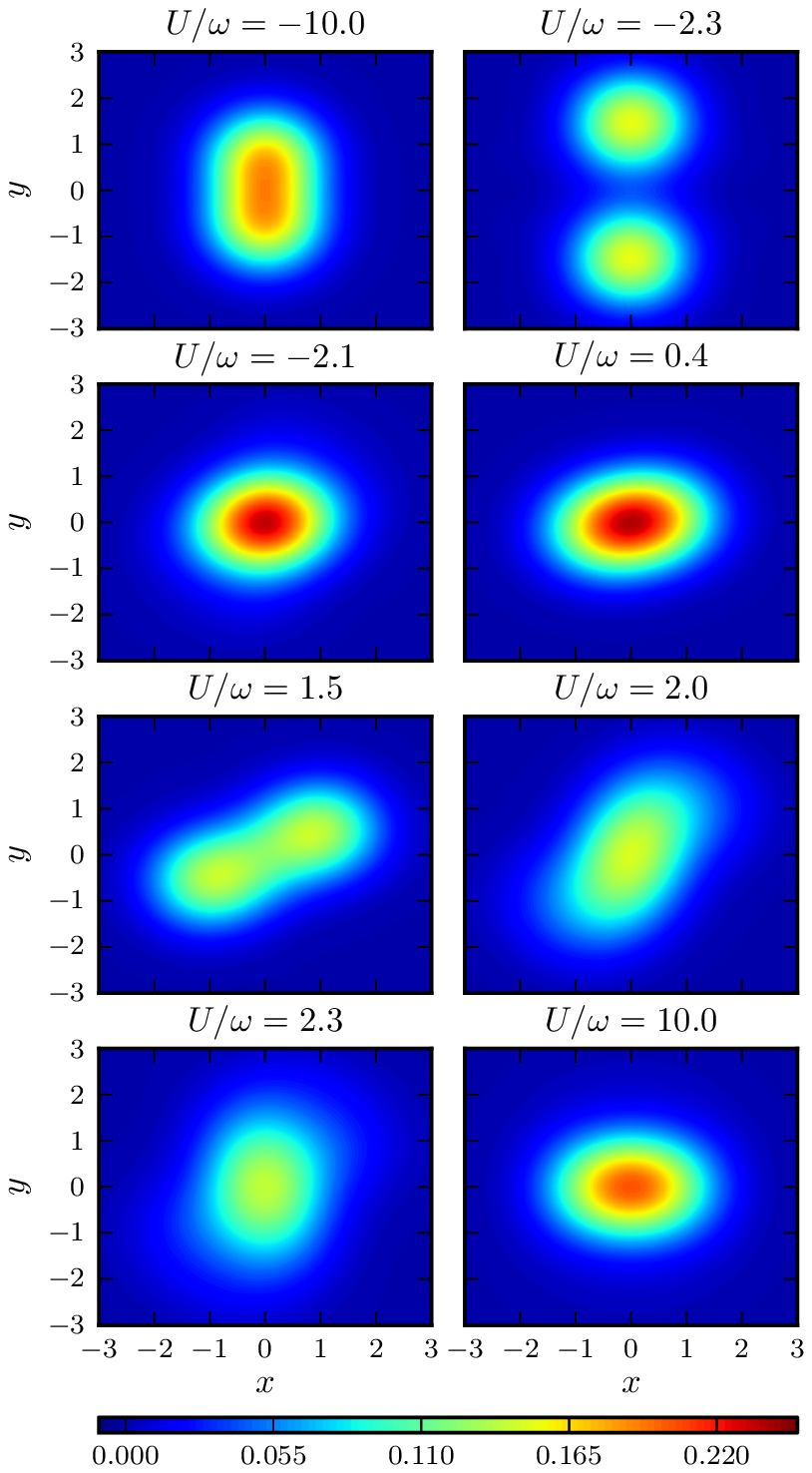}
  \caption{\label{fig:wigner_g0.1} Contour plots of the Wigner function $W(\alpha)$, with $\alpha = (x+iy)/\sqrt{2}$, for a series of values of $U$ along the middle horizontal cut in \fig{fig:phasedia}: $\{\omega_0,\omega,\kappa,g\} = \{0.05,1.0,0.2,0.1\} 2\pi\times\text{MHz}$, $N=50$.}
\end{figure}
\begin{figure}
  \centering
  \includegraphics{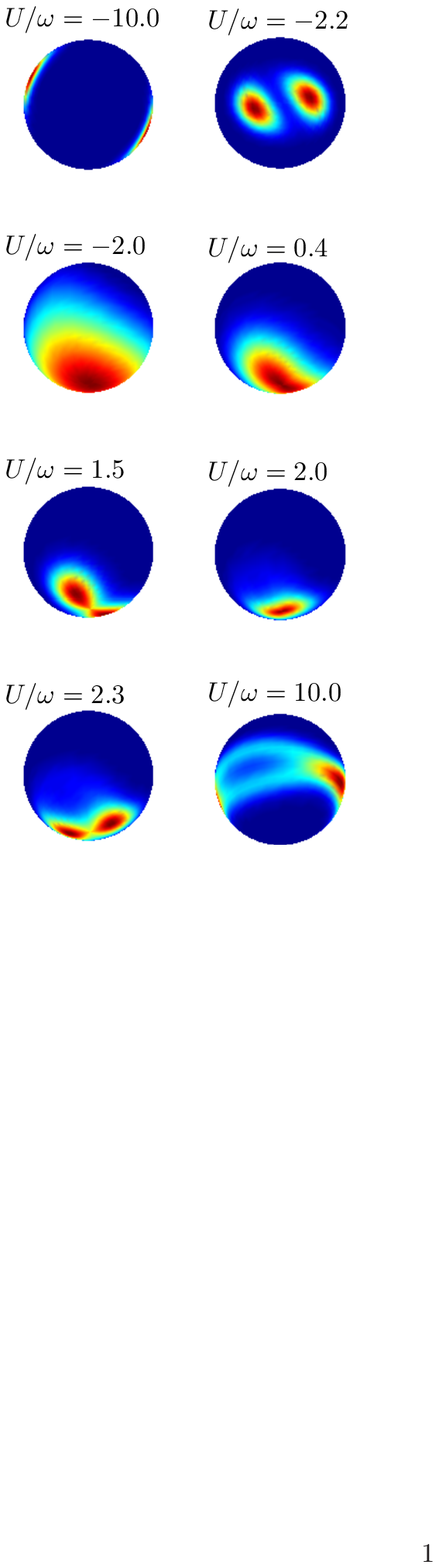}
  \caption{\label{fig:qfunc_g0.1} Steady state atomic $Q$-function, $Q(\theta,\phi)$, plotted on the Bloch sphere, for a series of values of $U$ along the middle horizontal cut in \fig{fig:phasedia}: $\{\omega_0,\omega,\kappa,g\} = \{0.05,1.0,0.2,0.1\} 2\pi\times\text{MHz}$, $N=50$. The top two spheres are rotated $180^\circ$ around the z-axis to better display their two-peak structure. The colour scheme goes from blue to red, indicating low to high concentration, respectively. The individual subplots are not on the same color scale.}
\end{figure}

Having studied the behaviour at low $g$, and seen some of the non-trivial finite-$N$ effects that appear, we now move on to study the superradiant phases that can appear for larger $g$. In \fig{fig:output_g0.1} we plot steady state expectation values as $U$ is varied along $g/(2\pi)=0.1$ MHz, i.e., the middle dashed line in \fig{fig:phasedia}. We compare three different atom numbers, $N=10$, $30$, and $50$, and observe an increasing agreement with the semiclassical solutions as $N$ grows. In particular, the results clearly point towards a first-order phase transition from a superradiant phase to the normal phase at $U\simeq-2.0\omega$. Comparing with the semiclassical predictions, we identify the phase for $U <-2\omega$ as the SRB phase. Less clear is the second order phase transition from the normal to the SRA phase around $U \simeq 0.5\omega$. The semiclassical analysis also predicts a narrow normal phase region around $1.9\omega < U < 2.0\omega$. For the finite $N$ results the only evidence of this region is a slight dip in the photon number. This is perhaps not surprising given how narrow in parameter space this region is, and given the influence of quantum fluctuations.  
Then, finally, as $U$ increases past $U = 2.0\omega$, one enters the semiclassical limit cycle region, as in the low $g$ case considered above.

In \fig{fig:wigner_g0.1} we show contour plots of the steady state Wigner function for the cavity mode, with atom number $N=50$. We focus on values close to where the various phase transitions occur. At large negative $U$, we are in the superradiant phase, where the Wigner function has a two-peak structure. The two-peak structure becomes more clear as $U$ is increased towards $U=-2\omega$. Close to this value, the Wigner function rapidly changes into a single-peaked function centered at the origin, indicating the normal phase. The Wigner function starts splitting again as $U$ crosses over beyond $U=0.5\omega$. The splitting is quite gradual, but one can see the Wigner function starting to deform around $U=0.4\omega$. At $U=1.5\omega$, the two-peak structure characteristic of a superradiant phase is clearly visible. At around $U=2.0\omega$, the Wigner function changes into an elongated shape around the origin, which we identify with the limit cycle region. In \fig{fig:qfunc_g0.1} we show the atomic $Q$-function at the same values of $U$. There is evidently a close correspondence between changes in the atomic state and changes in the state of the cavity field.

\subsection{$g/(2\pi) = 0.17$ {\rm MHz}}

\begin{figure}
  \centering
  \includegraphics{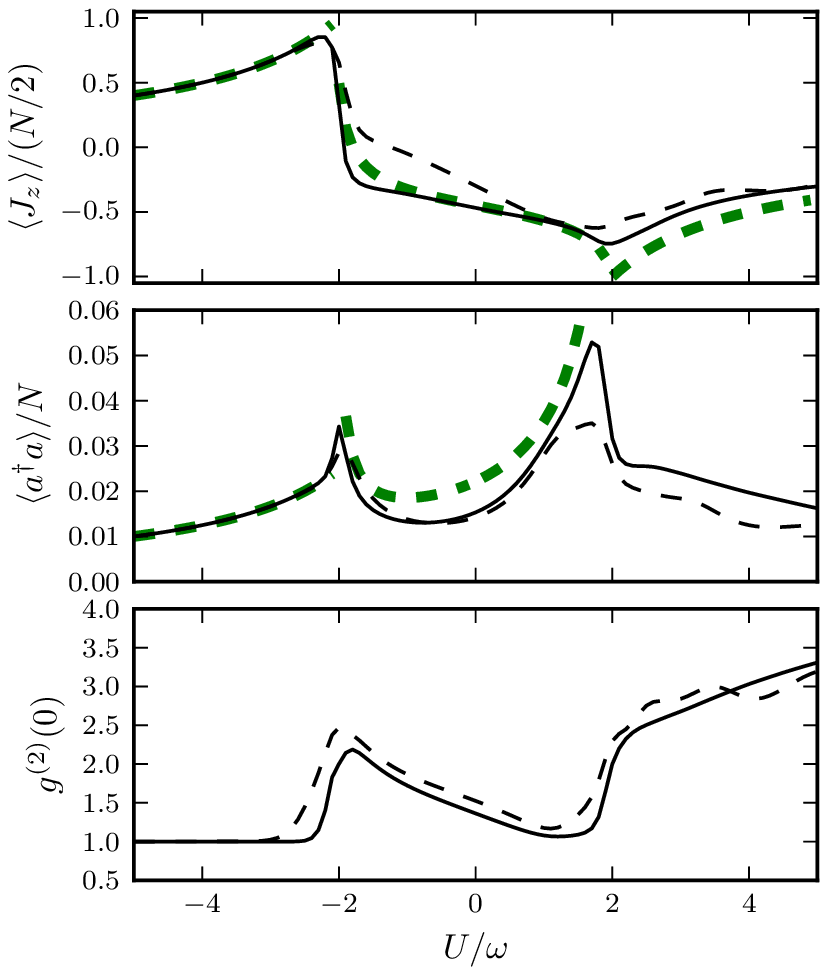}
  \caption{\label{fig:output_g0.17}  Steady state expectation values, $\braket{J_z}$, $\braket{a^\dag a}$ and $g^{(2)}(0)$, as $U$ is varied along the upper horizontal cut in \fig{fig:phasedia}: $\{\omega_0,\omega,\kappa,g\} = \{0.05,1.0,0.2,0.17\}\,2\pi\times\text{MHz}$: $N=10$ (dotted), $N=30$ (solid). Semiclassical solutions are included as dashed green lines for comparison.}
\end{figure}
\begin{figure}
  \centering
  \includegraphics{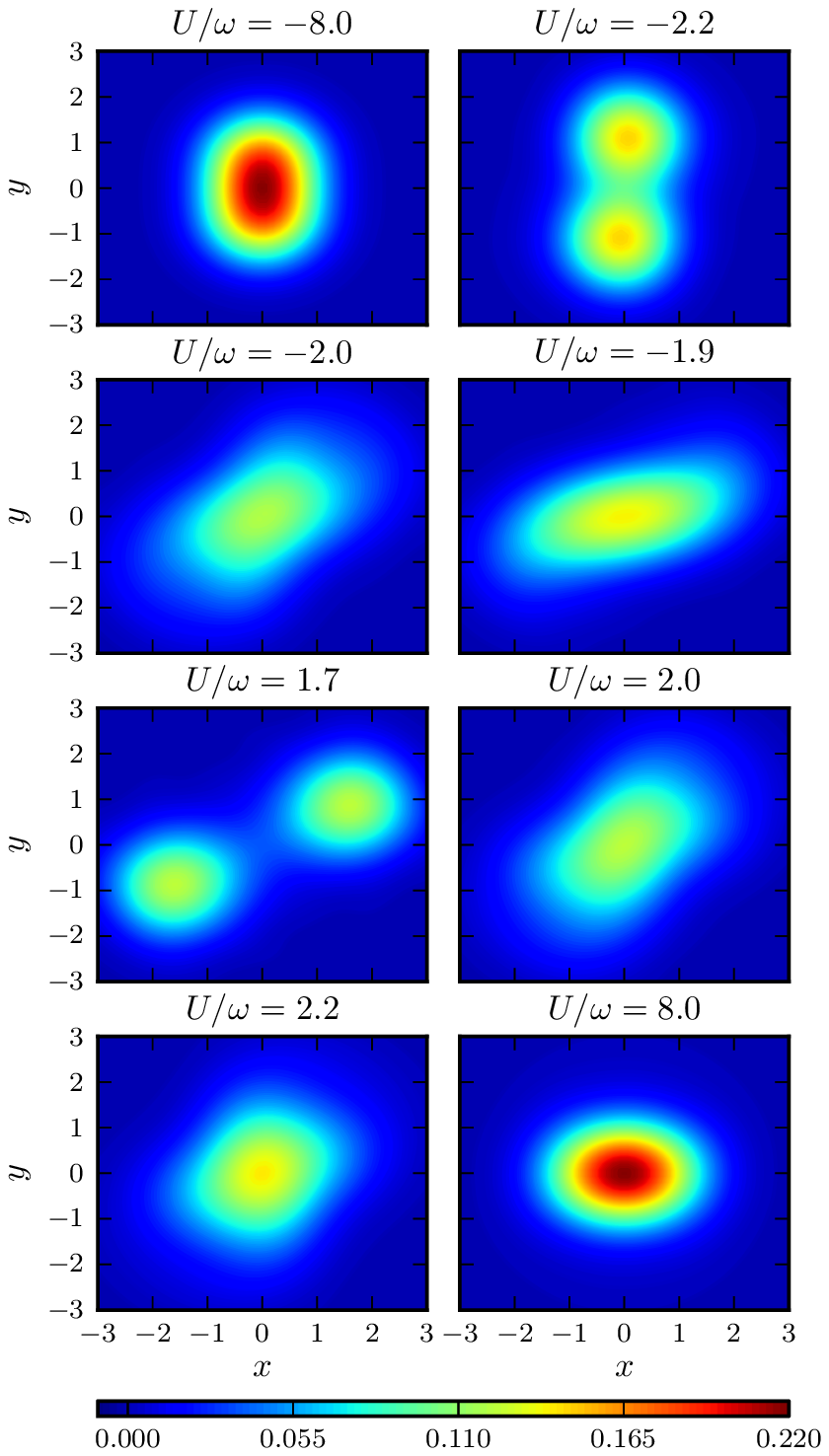}
  \caption{\label{fig:wigner_g0.17} Contour plots of the Wigner function $W(\alpha)$, with $\alpha = (x+iy)/\sqrt{2}$, for a series of values of $U$ along the upper horizontal cut in \fig{fig:phasedia}: $\{\omega_0,\omega,\kappa,g\} = \{0.05,1.0,0.2,0.17\}\,2\pi\times\text{MHz}$, $N=30$.}
\end{figure}
\begin{figure}
  \centering
  \includegraphics{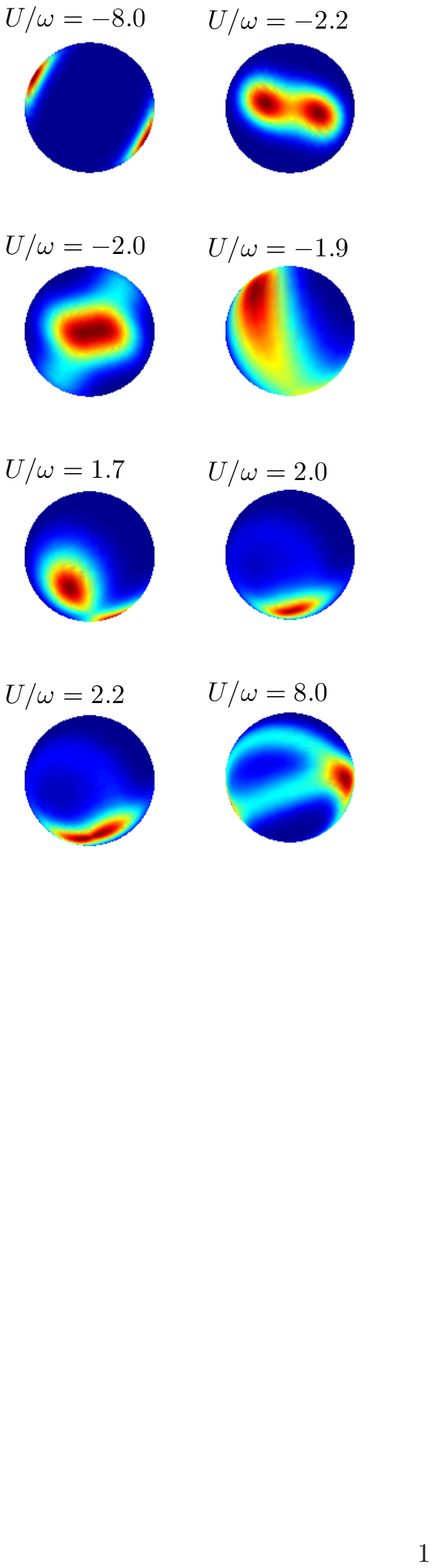}
  \caption{\label{fig:qfunc_g0.17} Steady state atomic $Q$-function, $Q(\theta,\phi)$, plotted on the Bloch sphere, for a series of values of $U$ along the upper horizontal cut in \fig{fig:phasedia}: $\{\omega_0,\omega,\kappa,g\} = \{0.05,1.0,0.2,0.17\}\,2\pi\times\text{MHz}$, $N=30$. The first three spheres (starting from the top, traversing rows first) are rotated $180^\circ$ around the z-axis. The colour scheme goes from blue to red, indicating low to high concentration, respectively. The individual subplots are not on the same color scale.}
\end{figure}

Finally, we consider an even larger value for $g$. The top dashed line in \fig{fig:phasedia} is at $g/(2\pi) = 0.17$ MHz. We show the variation in steady state expectation values, the Wigner function, and the atomic $Q$-function in \fig{fig:output_g0.17}, \fig{fig:wigner_g0.17} and \fig{fig:qfunc_g0.17}, respectively. The results are quite similar to the results for $g/(2\pi) = 0.1$ MHz in the previous section, the main difference being that the transition at $U=-2.0\omega$ is from the SRB phase to the SRA phase. In the Wigner function, this shows up as a rotation of the two peak structure in the $x$-$y$ plane. The expectation values show reasonably good agreement with the semiclassical solutions even for $N=30$.

\subsection{\label{sec:corrspectra}Steady state field correlation functions and power spectra}

\begin{figure}
  \centering
  \includegraphics{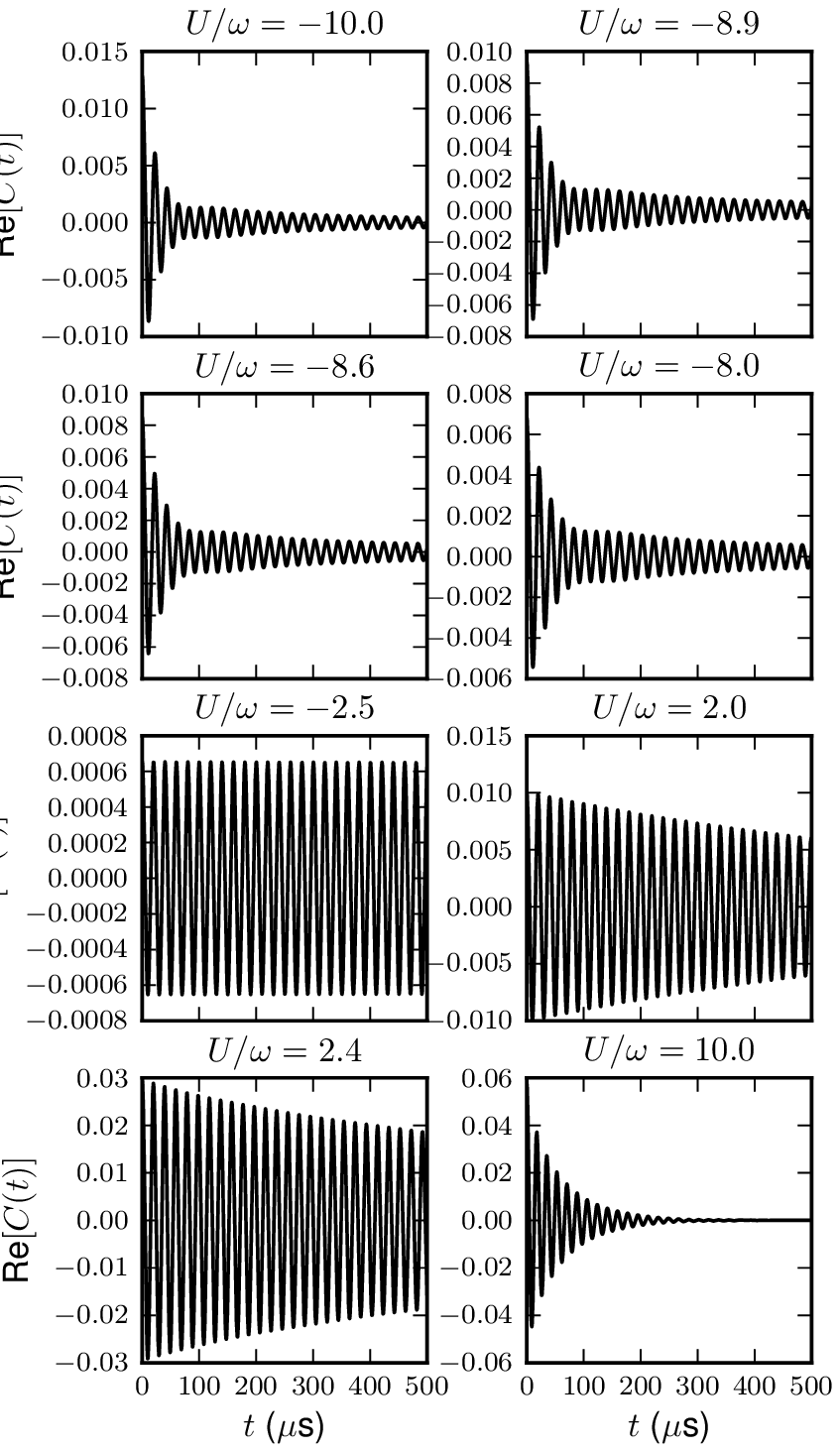}
  \caption{\label{fig:corr_g0.01} Field correlation function, $C(t)$, for a series of $U$ along the lower horizontal cut of \fig{fig:phasedia}: $\{\omega_0,\omega,\kappa,g\} = \{0.05,1.0,0.2,0.01\}\,2\pi\times\text{MHz}$, $N=90$.}
\end{figure}

\begin{figure}
  \centering
  \includegraphics{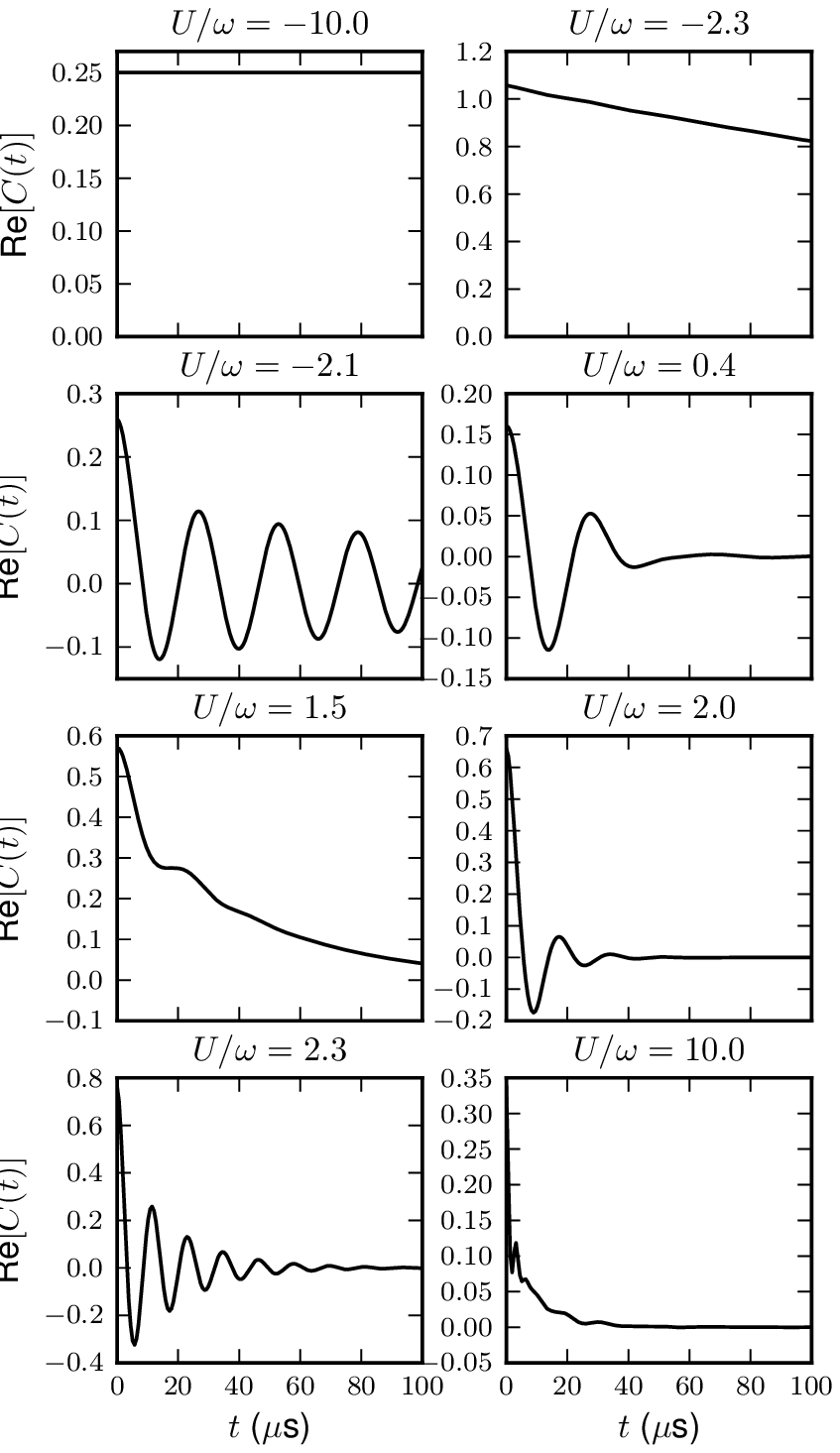}
  \caption{\label{fig:corr_g0.1} Field correlation function, $C(t)$, for a series of $U$ along the middle horizontal cut of \fig{fig:phasedia}: $\{\omega_0,\omega,\kappa,g\} = \{0.05,1.0,0.2,0.1\}\,2\pi\times\text{MHz}$, $N=50$.}
\end{figure}

\begin{figure}
  \centering
  \includegraphics{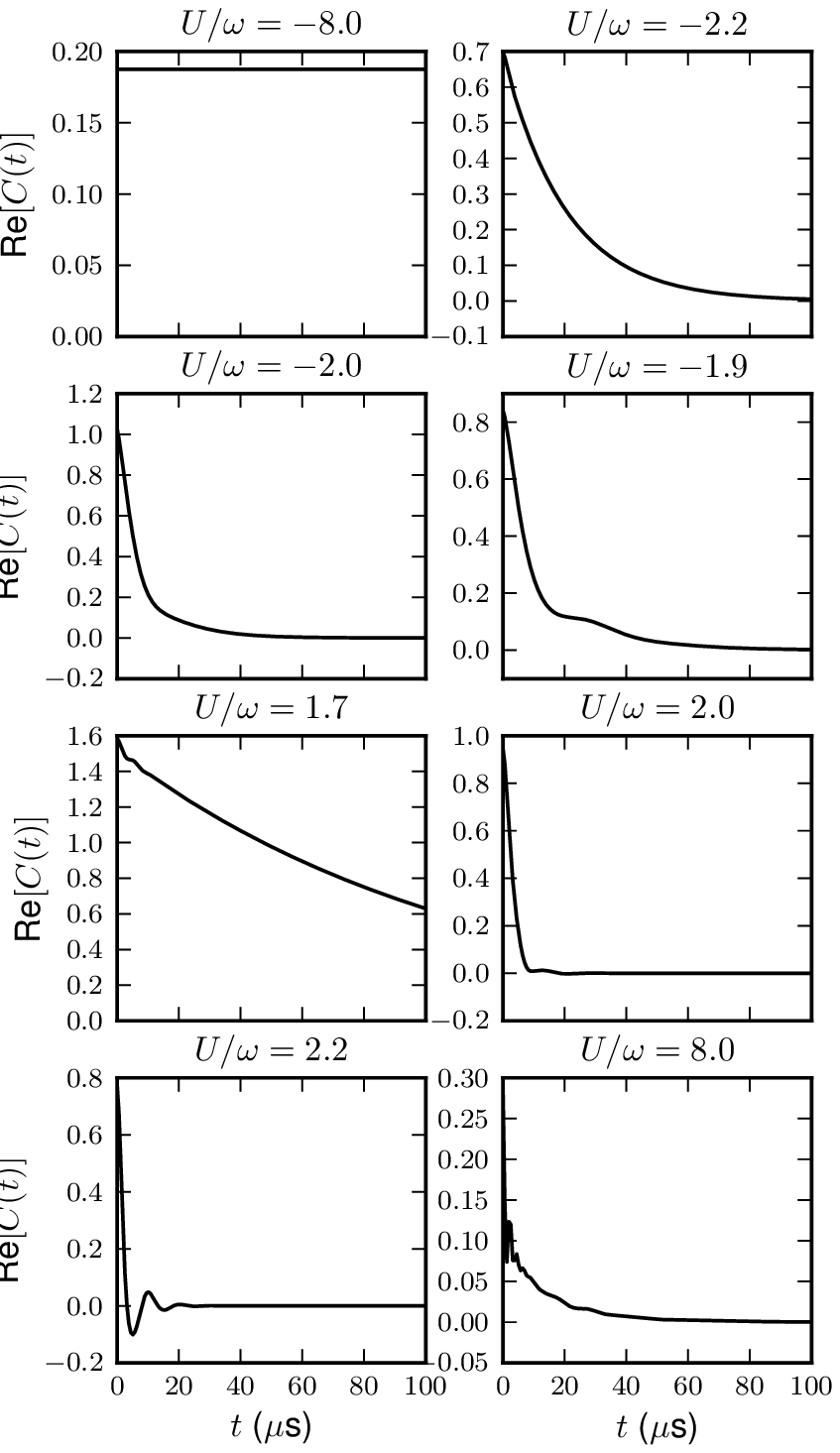}
  \caption{\label{fig:corr_g0.17} Field correlation function, $C(t)$, for a series of $U$ along the upper horizontal cut of \fig{fig:phasedia}: $\{\omega_0,\omega,\kappa,g\} = \{0.05,1.0,0.2,0.17\}\,2\pi\times\text{MHz}$, $N=30$}
\end{figure}

An aspect, and indeed a concern, of some experimental significance is the large variation in the characteristic timescales of the evolution in different regions of parameter space. The consequence of this for finite time experiments has been noted and discussed in some detail in \cite{Keeling10,Bhaseen12}. Here, we examine the characteristic timescales for the parameter ranges considered in the previous section by calculating the two-time amplitude correlation function for the cavity field, defined by
\begin{eqnarray*}
  C(t) := \braket{a\dagg(t) a(0)} - |\braket{a}_\text{ss}|^2,
\end{eqnarray*}
where the steady state coherent intensity, $|\braket{a}_\text{ss}|^2$, is subtracted for convenience. In figures \ref{fig:corr_g0.01}--\ref{fig:corr_g0.17}, we show the cavity field correlation function for the three different values of $g$ previously considered, $g/(2\pi) = \{0.01, 0.1, 0.17\}$ MHz, and for a series of values of $U$ corresponding to those investigated earlier in the Wigner and $Q$-function plots.

The correlation function plots illustrate clearly the wide range of dynamical time scales produced in the system through variation of the nonlinear coupling strength $U$, but we note in particular the exceedingly long-lived correlations for some parameter sets; in particular, for large negative $U$ and the two larger values of $g$ (i.e., $U/\omega = -10.0$ in \fig{fig:corr_g0.1} and $U/\omega = -8.0$ in \fig{fig:corr_g0.17}). The decay is not visible on the scale shown in the figures. These long-lived correlations can be related to the well-separated peaks in the atomic $Q$-functions shown for the corresponding parameters in \fig{fig:qfunc_g0.1} and \fig{fig:qfunc_g0.17}, respectively, and, in particular, very weak coupling (or rate of ``tunneling'') between the two distinct states associated with these peaks (note that the cavity field is close to the vacuum state for these values of $U$). We also note long time scales involved close to the phase boundaries, $U/\omega \simeq \pm 2.0$, characteristic of critical slowing down. 

\begin{figure}
  \centering
  \includegraphics{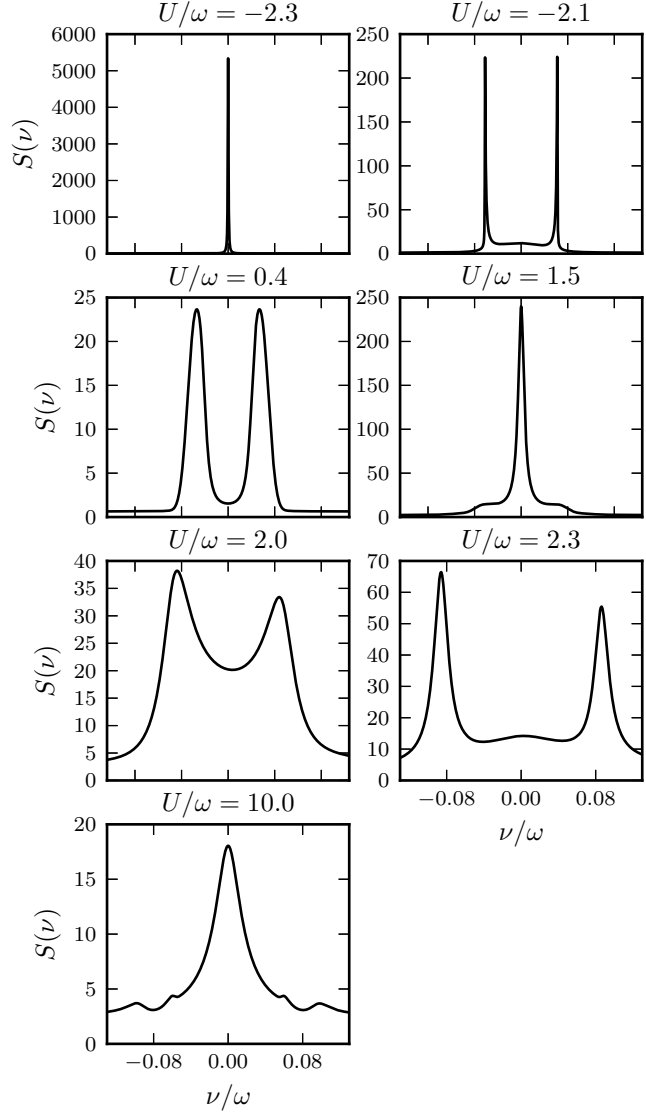}
  \caption{\label{fig:spectrum_g0.1} Power spectra, $S(\omega)$, for a series of $U$ along the middle horizontal cut of \fig{fig:phasedia}: $\{\omega_0,\omega,\kappa,g\} = \{0.05,1.0,0.2,0.1\}\,2\pi\times\text{MHz}$, $N=50$.}
\end{figure}

\begin{figure}
  \centering
  \includegraphics{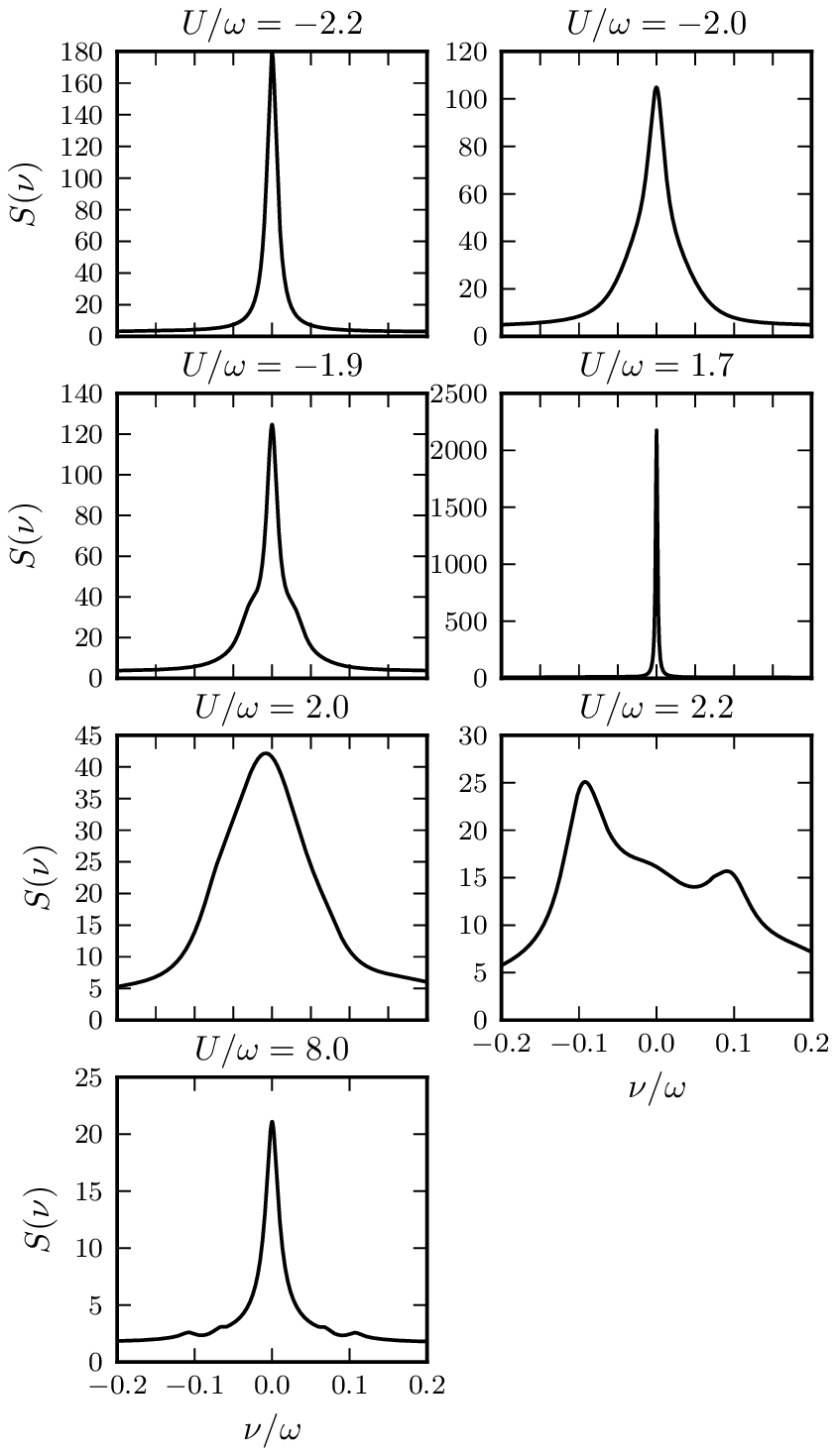}
  \caption{\label{fig:spectrum_g0.17} Power spectra, $S(\omega)$, for a series of $U$ along the upper horizontal cut of \fig{fig:phasedia}: $\{\omega_0,\omega,\kappa,g\} = \{0.05,1.0,0.2,0.17\}\,2\pi\times\text{MHz}$, $N=30$.}
\end{figure}

The Fourier transform of the correlation function,
\begin{eqnarray*}
  S(\nu) := \int_{-\infty}^{\infty} \dx{t} C(t)\exp(-i\nu t) ,
\end{eqnarray*}
gives the power spectrum of the cavity (output) field. 
We compute the spectrum numerically using a fast Fourier transform of the discrete time series shown in figures \ref{fig:corr_g0.01}--\ref{fig:corr_g0.17}. This can be problematic due to the slow decay of the correlation function for some parameter sets, and we do not include spectra for those cases where the field-correlations have not decayed for times on the order of a few ms. In \fig{fig:spectrum_g0.1} and \fig{fig:spectrum_g0.17} we show power spectra for the $g/(2\pi) = \{ 0.1, 0.17\}$ MHz cases, respectively. The power spectra provide an alternative (and directly measurable) means of studying the structure and dynamics of the system as $U$ is varied; for example, very slow time scales in the dynamics are manifest as extremely sharp and pronounced spectral features.

\subsection{Field-spin entanglement}

\begin{figure}
  \centering
  \includegraphics{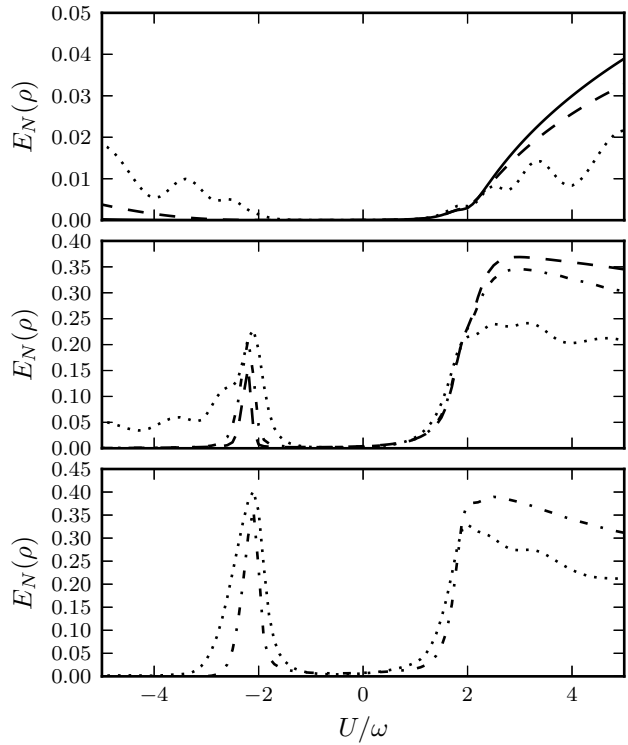}
  \caption{\label{fig:logneg} Logarithmic negativity in the steady state as $U$ is varied along the horizontal cuts of \fig{fig:phasedia}. Top panel: $\{\omega_0,\omega,\kappa,g\} = \{0.05,1.0,0.2,0.01\}\,2\pi\times\text{MHz}$, $N=10$ (dotted), $N=50$ (dashed) and $N=90$ (solid). Middle panel: $\{\omega_0,\omega,\kappa,g\} = \{0.05,1.0,0.2,0.1\}\,2\pi\times\text{MHz}$, $N=10$ (dotted), $N=30$ (dash-dotted) and $N=50$ (dashed). Bottom panel: $\{\omega_0,\omega,\kappa,g\} = \{0.05,1.0,0.2,0.17\}\,2\pi\times\text{MHz}$, $N=10$ (dotted) and $N=30$ (dash-dotted).}
\end{figure} 

Bipartite entanglement is believed to play a central role in quantum phase transitions, in the manner of divergence of correlation length at a critical point \cite{Osterloh02,Vidal03,Vidal04}. This has been studied in the Dicke model in particular, both in terms of the bipartite entanglement between the cavity field and the collective spin, and in terms of entanglement between individual spins \cite{Lambert04,Lambert05,Dimer07}.

Here we use the logarithmic negativity as a measure for the entanglement between the cavity mode and the collective atomic spin. This measure puts an upper bound on the amount of distillable entanglement between the two parts \cite{Vidal02}. The logarithmic negativity is given by
\begin{eqnarray}
  E_N(\rho) = \log_2||\rho^{\Gamma_A}||_1
\end{eqnarray}
where $\rho^{\Gamma_A}$ refers to the partial transpose of $\rho$ over the subsystem $A$, and $||X||_1 = \tr\sqrt{X\dagg X}$ is the trace norm of $X$. In \fig{fig:logneg} we show the logarithmic negativity in the steady state, as a function of $U$, for the three different cuts of the phase diagram (\fig{fig:phasedia}) considered above: $g/(2\pi) = \{0.01$, $0.1$, $0.17\}$ MHz.

The results indicate a low degree of distillable entanglement for large $N$, except close to the superradiant phase transition around $U \simeq -2\omega$ (assuming $g$ is large enough), in the SRA phase close to $U=2\omega$, and, interestingly, in the semiclassical limit cycle region, where $U > 2.0\omega$. For low $g$, the results add to what was observed in the expectation values considered earlier, namely a second order phase transition at $U=2\omega$, whereas the change in stability of the inverted phase at $U=-2\omega$, that was predicted by the semiclassical analysis, is not related to a phase transition for the quantum system. For higher $g$, however, the results point towards a first order phase transition at this value of $U$.

\section{Large $|U|$ behaviour}

\begin{figure}
  \centering
  \includegraphics{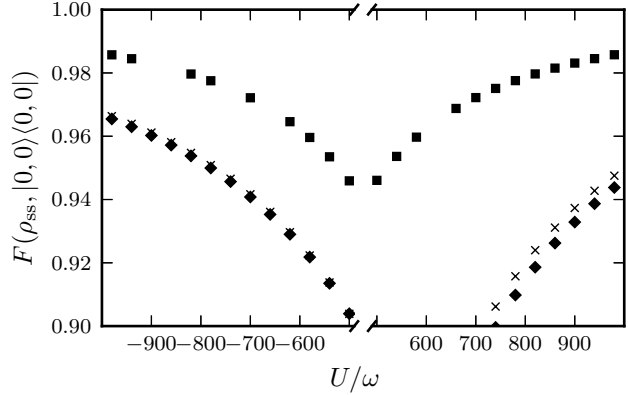}
  \caption{\label{fig:fidelity_largeU} Fidelity with the state $\ket{0,0}$, for the three different cuts of the phase diagram: $\{\omega_0,\omega,\kappa,g\} = \{0.05,1.0,0.2,0.01\}\,2\pi\times\text{MHz}$, $N=90$ (squares), $\{\omega_0,\omega,\kappa,g\} = \{0.05,1.0,0.2,0.1\}\,2\pi\times\text{MHz}$, $N=50$ (crosses) and  $\{\omega_0,\omega,\kappa,g\} = \{0.05,1.0,0.2,0.17\}\,2\pi\times\text{MHz}$, $N=30$ (diamonds).}
\end{figure}

The results so far point towards some similarities between large negative and positive $U$, especially for low $g$. The cavity photon number tends towards zero as $|U|$ grows, as does $\braket{J_z}$. Further investigation shows that for large $|U|$, and finite $N$, the atomic $Q$-function develops a distinctly ring-type structure, which moves towards the equator of the Bloch sphere with growing $|U|$, pointing towards a well-defined value for $J_z$. Calculation of the purity of the steady state, $\tr[\rho_\text{ss}^2]$, further reveals the approach towards a pure state with growing $|U|$ and in turn towards a simplified dynamic for large $|U|$. In fact, investigation shows that the steady state approaches the state $\ket{0}_\text{cav}\otimes\ket{J_z=0}$ in this limit, where $\ket{J_z = j_z}$ is a $J_z$ eigenstate with eigenvalue $j_z$. 
%In \fig{fig:qfunc_largeU} we show plots of the atomic q-function at large values for $|U|$, for $g/(2\pi) = 0.01$ and $0.1$ MHz, displaying the aforementioned ``ring structure''. 
In \fig{fig:fidelity_largeU} we show the fidelity of the steady state with the state $\ket{0,0} \equiv \ket{0}_\text{cav}\otimes\ket{J_z = 0}$ for $500$ MHz $<|U|/(2\pi)<$ $1000$ MHz, where the fidelity of two quantum states $\rho$ and $\sigma$ is defined as
\begin{eqnarray}
  F(\rho,\sigma) = \tr\left[\sqrt{\sqrt{\rho}\sigma\sqrt{\rho}}\right].
\end{eqnarray}

%\alg{Could also add example q-function plots for large $|U|$ values \dots}

\section{Conclusions}

We have studied, by numerical means, a generalized Dicke model, where the long time attractor for the model is a dynamical quantum steady state, resulting from a balance between photon production, due to the counter-rotating terms of the Dicke model Hamiltonian, and decay of the cavity field. This generalized model has recently received attention due to its realization in a quantum gas experiment using an optical cavity. Other recent experimental demonstrations of self-organization for a large number of rubidium atoms coupled to a high-finesse cavity also point towards a promising potential realization of the model.

We have in particular focused on the consequences of a non-linear coupling between the cavity photon number and the collective atomic inversion, written as $U J_z a\dagg a/N$, that is not present in the conventional Dicke model. If the non-linear coupling constant, $U$, becomes comparable to the cavity frequency, $\omega$, it can qualitatively change the steady state of the system. We have investigated this in detail, including quantum fluctuations and for a finite number of atoms, and seen how the steady state can be taken through a number of distinct phase transitions, which are observable, for example, as sharp changes in the cavity output field, as $U$ is varied. The predictions of an approximate, semiclassical model are seen to be approached as the number of atoms is increased.

For large values of the non-linear coupling, the atomic steady state approaches a state with a well-defined value for the inversion, $J_z$, whereas $J_{\{x,y\}}$ become delocalized---the latter corresponding to periodic orbits for the expectation values in the semiclassical approximation. In particular, as $U$ tends to infinity, the steady state approaches the pure state $\ket{0}_\text{cav} \otimes \ket{J_z=0}_\text{spin}$, i.e., the vacuum state for the cavity, while the atoms are in the maximally entangled $J_z=0$ eigenstate. For a finite number of atoms, $N$, this is also the case for $U\to-\infty$, whereas in the thermodynamic limit, $N\to\infty$, a mixture of $J_z = \pm N/2$, eigenstates is predicted for large negative $U$. For finite $N$, this co-existence phase is observed for negative $U < -2\omega$, but a sharp change in the atomic state appears at $U \simeq -2\omega_0 N$, corresponding to the development of a ring-type structure in the atomic $Q$-function, which eventually approaches the $J_z=0$ eigenstate, as in the $U\to\infty$ case.

The rich behaviour of the system, due to the presence of cavity decay and the non-linear atom-photon coupling, together with the readily observable cavity output field, suggests that various experimental implementations might offer observations of not only the conventional superradiant Dicke quantum phase transition, but also other novel quantum phases and phase-boundaries as well. We hope that this work might further stimulate experimental, as well as theoretical, investigations of the model.

% iop acknowledgment style
\ack

The authors wish to acknowledge the contribution of NeSI high-performance computing facilities to the results of this research. New Zealand's national facilities are provided by the NZ eScience Infrastructure and funded jointly by NeSI's collaborator institutions and through the Ministry of Science \& Innovation's Research Infrastructure programme. URL http://www.nesi.org.nz. The authors thank Howard Carmichael for helpful discussions.

%\bibliography{arne}

\end{document}